\documentclass[%
 reprint,
 superscriptaddress,
 amsmath,amssymb,
 aps,
prb,
]{revtex4-2}

\makeatletter
\def\@hangfrom@section#1#2#3{\@hangfrom{#1#2}#3}%
\def\@hangfroms@section#1#2{#1#2}%
\makeatother

\usepackage[dvipsnames]{xcolor}%

\usepackage{physics}%
\usepackage{graphicx}%
\usepackage{dcolumn}%
\usepackage{bm}%

\usepackage{float}
\usepackage{booktabs}
\usepackage{multirow}
\usepackage{placeins}

\usepackage{xspace}

\begin{document}

\preprint{Draft}

\title{Modelling complex proton transport phenomena - Exploring the limits of fine-tuning and transferability of foundational machine-learned force fields}

\author{Malte Grunert}
\affiliation{These authors contributed equally to this work.}
\affiliation{Theoretical Physics I, Institute of Physics, Technische Universit\"at Ilmenau, 98693 Ilmenau, Germany}
\affiliation{Center of Micro- and Nanotechnologies, Technische Universit\"at Ilmenau, 98693 Ilmenau, Germany}

\author{Max Großmann}
\affiliation{These authors contributed equally to this work.}
\affiliation{Theoretical Physics I, Institute of Physics, Technische Universit\"at Ilmenau, 98693 Ilmenau, Germany}
\affiliation{Center of Micro- and Nanotechnologies, Technische Universit\"at Ilmenau, 98693 Ilmenau, Germany}

\author{Jonas Hänseroth}
\affiliation{Theoretical Solid State Physics, Institute of Physics, Technische Universität Ilmenau, 98693 Ilmenau, Germany}

\author{Aaron Flötotto}
\affiliation{Theoretical Solid State Physics, Institute of Physics, Technische Universität Ilmenau, 98693 Ilmenau, Germany}

\author{Jules Oumard}
\affiliation{Theoretical Solid State Physics, Institute of Physics, Technische Universität Ilmenau, 98693 Ilmenau, Germany}

\author{Johannes Laurenz Wolf}
\affiliation{Theoretical Solid State Physics, Institute of Physics, Technische Universität Ilmenau, 98693 Ilmenau, Germany}

\author{Erich Runge}
\affiliation{Theoretical Physics I, Institute of Physics, Technische Universit\"at Ilmenau, 98693 Ilmenau, Germany}
\affiliation{Center of Micro- and Nanotechnologies, Technische Universit\"at Ilmenau, 98693 Ilmenau, Germany}

\author{Christian Dreßler}
\affiliation{Theoretical Solid State Physics, Institute of Physics, Technische Universität Ilmenau, 98693 Ilmenau, Germany}
\email{christian.dressler@tu-ilmenau.de}

\date{\today}

\begin{abstract}
The solid acids CsH$_2$PO$_4$ and Cs$_7$(H$_4$PO$_4$)(H$_2$PO$_4$)$_8$ pose significant challenges for the simulation of proton transport phenomena. 
In this work, we use the recently developed machine-learned force field (MLFF) MACE to model the proton dynamics on nanosecond time scales for these systems and compare its performance with long-term \textit{ab initio} molecular dynamics (AIMD) simulations.
The MACE-MP-0 foundation model shows remarkable performance for all observables derived from molecular dynamics (MD) simulations, but minor quantitative discrepancies remain compared to the AIMD reference data.
However, we show that minimal fine-tuning - fitting to as little as 1~ps of AIMD data - leads to full quantitative agreement between the radial distribution functions of MACE force field and AIMD simulations.
In addition, we show that traditional long-term AIMD simulations fail to capture the correct qualitative trends in diffusion coefficients and activation energies for these solid acids due to the limited accessible time scale.
In contrast, accurate and convergent diffusion coefficients can be reliably obtained through multi-nanosecond long MD simulations using machine-learned force fields.
The obtained qualitative and quantitative behavior of the converged diffusion coefficients and activation energies now matches the experimental trends for both solid acids, in contrast to previous AIMD simulations that yielded a qualitatively wrong picture.
\end{abstract}

\maketitle

\section*{Introduction}

The unsupervised, black-box-like prediction of atomic forces with \textit{ab initio} accuracy has been a central goal of the MLFF paradigm ever since Behler and Parrinello published their seminal paper on the representation of potential-energy surfaces with neural networks in 2007 \cite{behler2007}.
Several generations of machine learning approaches, such as feed-forward and graph neural networks, and kernel regression methods, e.g., Gaussian Approximation Potentials (GAP) \cite{bartok2010}, have been successfully applied to molecular dynamics simulations \cite{unke2021, friederich2021, reiser2022}.
Each of these methods relies on a descriptor that represents the local atomic environment. 
A variety of different descriptors have been developed in the past, to name but a few: atom centered symmetry functions \cite{behler2007}, the smooth overlap of atomic positions descriptor \cite{bartok2010} and moment tensor potential basis functions \cite{shapeev2016}.

Inspired by the success of foundational models in natural language processing, recent efforts have proposed training MLFFs on large-scale datasets such as the Materials Project database \cite{Jain2013, Ong2015}. 
Foundational MLFFs can be thought of as fundamental models for atomistic materials chemistry that can generalize across different classes of materials while remaining easily adaptable to specific applications \cite{batia2023, Yang2024}.

These advances are driven by innovations in atomic environment descriptors, such as the atomic cluster expansion (ACE) \cite{drautz2019}, and the integration of equivariance principles as a core design element in graph neural networks \cite{reiser2022}.
The synergy between the ACE descriptor and the enforcement of equivariance has enabled the development of accurate and generalizable MLFFs \cite{Batatia2022, batzner2022, kovacs2023}, which can be considered foundation models.

In this paper, we focus on one of the most promising publicly available equivariant MLFFs \cite{batia2023}, and investigate its fine-tuning and transferability on the example of two particularly challenging compounds for simulating proton transport: the solid acids  CsH$_2$PO$_4$ (CDP) and Cs$_7$(H$_4$PO$_4$)(H$_2$PO$_4$)$_8$ (CPP).
We benchmark the performance of the MACE foundation model and evaluate the amount of fine-tuning required to achieve quantitative agreement between the radial distribution functions of MACE and \textit{ab initio} molecular dynamics (AIMD) simulations, as well as the degree of transferability of fine-tuned models to closely related material systems. 
We show how the limited time scales accessible in AIMD simulations prevent the calculation of diffusion coefficients that are even qualitatively consistent with existing experimental data.
Furthermore, we demonstrate that MLFF-based MD simulations on the nanosecond time scale yield converged diffusion coefficients that are in agreement with experiments.

\begin{figure*}[ht]
    \includegraphics{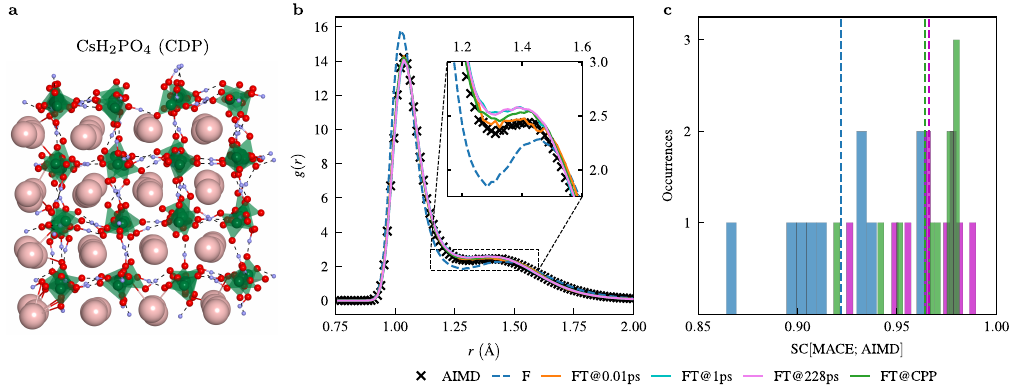}
    \caption{
    \label{fig:plot1}
    \textbf{a} Snapshot from an AIMD simulation of CDP. Color scheme for atoms: hydrogen in blue, oxygen in red, phosphorus in green, and caesium in pink. 
    \textbf{b} Comparison of the O-H radial distribution function $g(r)$ obtained from different MACE models (see main text) with $g(r)$ from an AIMD simulation. 
    The inset highlights the peak of $g(r)$ at $d_\mathrm{OH} = 1.5$~\AA, which is commonly referred to as the ”short strong hydrogen bond” or ”low/barrier hydrogen bond” \cite{bum2004, limbach2009, whiteley2017, dereka2021}. 
    \textbf{c} Histogram showing the similarity coefficients SC as defined by Eq.~(\ref{eqn:sc}) with respect to the AIMD results, i.e., $\mathrm{SC}[\mathrm{MACE};\,\mathrm{AIMD}]$ for all possible bond combinations for the three selected MACE models from \textbf{b}, see legend. 
    The average SC is indicated by vertical dashed lines in the corresponding color.
    }
\end{figure*}

\begin{figure*}[ht]
    \includegraphics{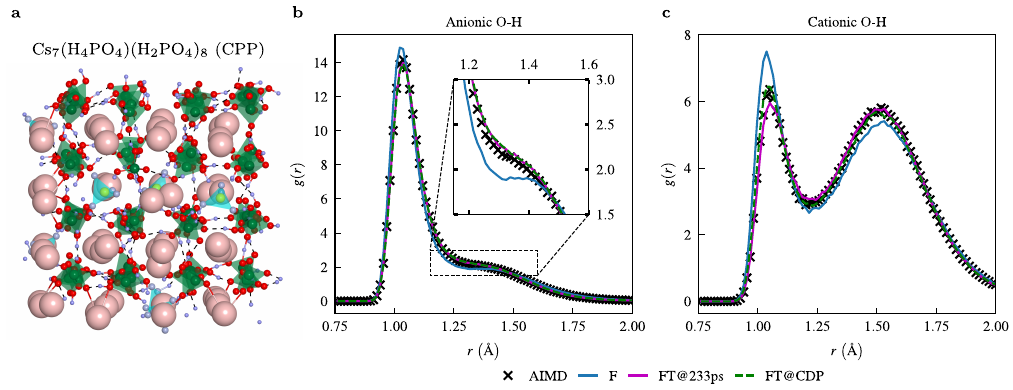}
    \caption{
    \label{fig:plot2}
    \textbf{a} Snapshot from an AIMD simulation of CPP. 
    Anionic H$_2$PO$_4^-$ tetrahedra are highlighted in green, and cationic H$_4$PO$_4^+$ tetrahedra are highlighted in yellow. 
    Color scheme for atoms: hydrogen in blue, oxygen in red, phosphorus in green/yellow (anionic/cationic), and caesium in pink. \textbf{b} Comparison of the anionic O-H radial distribution function $g(r)$ obtained from different MACE models (see main text) with $g(r)$ from an AIMD simulation. 
    The inset highlights the peak of $g(r)$ at $d_\mathrm{OH} = 1.5$~\AA, which is commonly referred to as the ”short-strong hydrogen bond” or ”low-barrier hydrogen bond”. 
    \textbf{c} Same as \textbf{b} but for the cationic O-H radial distribution function.
    }
\end{figure*}

Solid acids are inorganic crystalline compounds with molecular formulas similar to CsH$_\mathrm{y}$XO$_4$  ($\mathrm{X} = \mathrm{S}$, P, Se; $y = 1, 2$), where protons can act as charge carriers \cite{baranov03}.
These compounds undergo a superprotonic phase transition from a low- to a high-temperature phase, which is associated with a drastic increase in proton conductivity \cite{goni12}. 
While the low-temperature phases are almost proton insulators, the high-temperature phases are excellent proton conductors. 
The technological relevance of these materials arises from the fact that they enable water-free proton conduction, which qualifies them as suitable materials for proton exchange fuel cell membranes operating in the medium temperature range \cite{haile2001, chisholm09, haile07}.  
The most prominent representative of this class of compounds is CDP, which is already used as a membrane material in a commercially available fuel cell and exhibits the superprotonic phase transition (monoclinic to cubic phase) at 509~K \cite{boysen04, yamada04}.
A similar but chemically more intriguing solid acid, CPP, is derived from CDP by replacing one in eight caesium ions with the unusual tetrahydroxyphosphonium cation H$_4$PO$_4^+$, leading to the unique situation of anionic H$_2$PO$_4^-$ and cationic H$_4$PO$_4^+$ phosphate groups coexisting in a single crystalline phase \cite{wang20}. 
The coexistence was confirmed by $^{31}$P NMR and high-temperature X-ray crystal diffraction by the group of Sossina Haile~\cite{wang20}, who also synthesized this compound for the first time in 2020.
This finding is all the more remarkable because proton transfer between the phosphate groups is possible via a strong and fluctuating hydrogen bond network.
We have previously studied the distribution of protons in this compound using AIMD simulations \cite{dressler23} and were able to show that the H$_4$PO$_4^+$ cation introduces a complicated distribution of protons in CPP and that the different phosphate groups differ in their proton interaction. 
If only covalently bound protons are considered, the oxygen atoms of the "formally" cationic phosphate groups H$_4$PO$_4^+$ are bound to a smaller number of protons compared to the oxygen atoms of the anionic phosphate groups H$_2$PO$_4^-$. 
However, the "formal" protonation states can be maintained if both hydrogen and covalently bound hydrogen atoms are considered. 
All oxygen atoms in CPP are in a dynamic equilibrium (fluctuating on the sub-ps time scale) between hydrogen and covalently bonded states, which justifies counting the number of covalently and hydrogen bonded protons per oxygen atom followed by division of this number by two, as each proton is always involved in one hydrogen and one covalent bond.
Following this approach, the 'formal' protonation states H$_4$PO$_4^+$/H$_2$PO$_4^-$ are obtained.
The crystal structures of CDP and CPP, with the anionic H$_2$PO$_4^-$ and cationic H$_4$PO$_4^+$ phosphate groups highlighted by green and cyan tetrahedra, are shown in Fig.~\ref{fig:plot1}a and Fig.~\ref{fig:plot2}a, respectively. 

The proper description of these materials using MD simulations faces significant challenges.
As discussed in Ref.~\cite{dressler23} AIMD simulations are not able to predict the correct qualitative trend of the diffusion coefficients.  
Although the accuracy of the chosen level of theory (here: GGA-DFT) may be questionable, this discrepancy is most likely due to the short time scales accessible to AIMD simulations.
Classical force fields cannot accurately describe bond breaking and bond formation, which are fundamental steps in the proton conduction mechanism of this compound. 
In addition, classical force fields are unlikely to effectively capture the complex dynamics of these systems, where the anions and cations exhibit solid-like behavior, while the rotational dynamics of the phosphate groups are comparable to the liquid state.
Since the resulting fluctuating hydrogen bond network is the origin of the extraordinarily high proton conductivity in these compounds, its proper treatment is critical.
Promising alternatives to overcome these problems are multi-scale methods \cite{joerg2023}, which combine, for example, MD simulations with Monte Carlo approaches \cite{dressler2016,  kabbe2016, dressler2020a, dressler2020b}. Another approach is MLFFs, which aim to combine the speed of classical force fields with the precision of AIMD.
For the foundation model approach regarding MLFFs, another issue arises:
To our knowledge, the tetrahydroxphosphonium cation H$_4$PO$_4^+$ has been described only 5 times in the literature \cite{hibbert85, addison83, minkwitz99, mathew95, arlman37} and is therefore not or rarely included in any database used for the initial training of a foundation model. 
This leads to the four main questions of this work:
(a) How well do foundation models perform for solid acids?
(b) How much AIMD data is needed to fine-tune MLFFs for solid acids?
(c) How transferable is a fine-tuned model within its chemical class?
(d) Is an adequate description of the transport properties of these highly complex materials finally possible?

To investigate these questions, we have performed long-term AIMD and MLFF simulations using the MACE-MP-0 model with or without the empirical D3 dispersion correction \cite{Grimme-2010-ID679} (referred to as F and F+D3 in the following) and several fine-tuned versions of the MACE-MP-0 model for both compounds.
We fine-tune the MACE-MP-0 on different numbers of snapshots: every 10th snapshot for $0.01$~ps, $0.1$~ps and $1$~ps, and every 100th snapshot from a $228$~ps trajectory of CDP and $233$~ps trajectory of CPP. We name these models accordingly, i.e., FT@X with $\mathrm{X}\in \{0.01\mathrm{ps}, 0.1\mathrm{ps}, 1\mathrm{ps}, 228\mathrm{ps}/233\mathrm{ps}\}$.
This corresponds to $2$, $20$, $200$ and $458/466$ snapshots used for fine-tuning, respectively.
Note that to investigate the transferability of the MLFF models, we will apply the fully fine-tuned MACE model for CPP (FT@233ps) to molecular dynamics simulations of CDP, and vice versa. 
In the context of this transferability discussion, we refer to FT@228ps as FT@CDP and FT@233ps as FT@CPP.
The time step for all MD simulation was set to 0.5~fs.
Further details about the MD simulations and the fine-tuning process are provided in the Methods section.

\section*{Results}
The fine-tuned models exhibit very small force errors on the test set, e.g., around 35~meV\,{\AA}$^{-1}$ for the fully fine-tuned models. 
The test set was constructed by selecting equally spaced frames from the extended MD trajectories generated using the fully fine-tuned MACE models.
Forces and energies calculated for these frames via DFT were compared to those predicted by various MACE models. Detailed information on the test set errors are given in Supplementary~Note~3.
However, as was recently discussed in Ref.~\cite{fu22}, better accuracy in the forces does not always lead to accuracy in the trajectories and derived properties.
Therefore, we decide to evaluate the performance of the MLFF models by comparing radial distribution functions (RDF) from AIMD and MLFF simulations at 510~K.
This temperature was chosen because CDP undergoes its superprotonic phase transition at 509~K.
We calculated the RDFs for all possible atomic combinations found in CDP. 
The most important RDF in terms of predicting proton mobility, the OH-RDF, is shown in Fig.~\ref{fig:plot1}b.
The OH-RDF predicted by each MACE model, i.e., both fine-tuned and foundation models, are in excellent agreement with the RDF obtained from our AIMD simulation.   
Focusing on the region of the OH-RDF associated with "short-strong" or "low-barrier" hydrogen bonds ($d_\mathrm{OH}\in[1.25,,1.75]$~{\AA}) \cite{bum2004, limbach2009, whiteley2017, dereka2021}, notable differences between the predictions of the fine-tuned and foundation models become apparent (see inset in Fig.~\ref{fig:plot1}b).
This region of the OH-RDF is particularly critical for the accurate description of proton transport phenomena in solid acids, as proton jumps within these distances are characterized by low activation energies, making them highly important for the overall process. 
While the foundation model shows deviations from the AIMD reference data in this region, the fine-tuned models achieve a high degree of quantitative accuracy, effectively capturing the key features of the RDF.
Remarkably, the quantitative agreement of the fine-tuned MACE models is given for all levels of fine-tuning, i.e., even the fine-tuned models obtained by training on only one frame (FT@0.01ps) yield OH-RDFs with excellent accuracy.   
Comparing F and F+D3 in Supplementary~Fig.~1, we note that the inclusion of the D3 dispersion correction does not significantly change the RDFs. 
This is due to the fact that the intermolecular interactions of the present system are dominated by strong hydrogen bonds, for which dispersion forces do not play a major role.

To quantify the agreement between RDFs calculated from AIMD and MACE models, we use a similarity coefficient $\mathrm{SC}$ previously used in other contexts \cite{Biswas2023, Grunert2024}, which is defined as follows
\begin{equation}\label{eqn:sc}
    \mathrm{SC}\big[\tilde{g}(.);g(.)\big] = 1 - \frac{\int\big\vert \tilde{g}(r) - g(r)\big\vert\,dr}{\int \big\vert g(r)\big\vert\,dr}
\end{equation}
where $g(r)$, $\tilde{g}(r)$ are the radial distribution functions to be compared.
The $\mathrm{SC}$ value equals 1 for a perfect match between the functions being compared and decreases, potentially toward negative infinity, as the two functions become increasingly dissimilar.
The distribution of similarity coefficients and their mean value for three MACE models (F, FT@228ps, FT@CPP) are shown in Fig.~\ref{fig:plot1}c.
Even the worst similarity coefficients for each of the six variants of the MACE model considered are all above 0.85, the worst being $\mathrm{SC}=0.864$ for the CsP-RDF of the foundation model.
All fine-tuned models significantly outperform the foundation model, with all $\mathrm{SC}\geq0.92$.
The very good agreement between the RDFs obtained by AIMD and each variant of the MACE model can be seen in detail in Supplementary~Tab.~1.  

We repeat this procedure for CPP, where the longest AIMD trajectory used for fine-tuning is 233~ps long.
The OH-RDF between anionic (O$_\mathrm{an}$) and cationic (O$_\mathrm{cat}$) oxygen atoms and hydrogen atoms for CPP is shown in Fig.~\ref{fig:plot2}b and Fig.~\ref{fig:plot2}c, respectively.
Again, the foundation and fine-tuned MACE models are compared with the AIMD.
Just as with CDP, we find good qualitative agreement for the foundation model and excellent quantitative agreement for the fine-tuned model.
The general shape of the O$_\text{an}$H-RDF of CPP is similar to the shape of the OH-RDF of CDP. 
These RDFs are prototypes for OH-RDFs of excellent proton conductors, which form short and very strong hydrogen bonds.
In contrast to the O$_\text{an}$H-RDF, the O$_\text{cat}$H-RDF differs significantly. 
The area of the peak corresponding to the "short-strong" hydrogen bonds ($d_\mathrm{OH}\in[1.25,\,1.75]$~{\AA}) is larger than that of the covalently bonded hydrogen atoms ($d_\mathrm{OH}\in[0.9,\,1.2]$~{\AA}), i.e., more H atoms are bonded to the cationic oxygen atoms by a hydrogen bond than by a covalent bond. 
If we integrate the O$_\text{cat}$H-RDF, we find that there are a total of two protons bound to each oxygen atom, either by a hydrogen or a covalent bond. 
Thus, each oxygen atom of the cationic phosphate groups is involved in a bifurcated hydrogen bond. 
This remarkable protonation state is associated with the unusual tetrahydroxyphosphonium cation H$_4$PO$_4^{+}$, which, as mentioned above, is reported only about 5 times in the literature \cite{hibbert85, addison83, minkwitz99, mathew95, arlman37}.
Thus, the good qualitative agreement of the foundation model is therefore quite remarkable.

We recall that the atomistic structures of CDP and CPP are closely related.
They differ only in the substitution of one in every eight Cs$^+$ ions by the unusual tetrahydroxphosphonium cation. 
This observation motivated us to use the fully fine-tuned model for CDP (denoted as FT@CDP) to calculate the RDFs of CPP, see Fig.~\ref{fig:plot2}. 
The O$_\text{an}$H-RDFs and the O$_\text{cat}$H RDFs are in full agreement with the RDFs obtained from the fully fine-tuned model for CPP (FT@233ps) and also with the RDFs from the AIMD simulation.
The similarity coefficients for the RDFs obtained from all possible atomic combinations in  CPP are shown in Supplementary~Tab.~2, further underlining the results from the particular example of the OH-RDFs.
Applying the model fine-tuned on CPP (denoted as FT@CPP) to CDP works similarly well, see Fig.~\ref{fig:plot1}.
This is also a remarkable result: For the chemically closely related systems we have studied, fine-tuning on only one compound is required. 
The resulting model can be generalized to the other compound.

From our discussion of the RDFs of CDP and CPP obtained from AIMD and various MACE model simulations, we can now answer the first three questions posed above:
(1) Foundation models perform remarkably well even for exotic solid acids, but they do not provide full quantitative agreement, which may have a noticeable impact on derived properties.
(2) Fine-tuning to quantitative agreement is possible with a small number of AIMD snapshots, in extreme cases even a single snapshot, although increasing the number of snapshots improves quantitative agreement.
This raises interesting questions about the future need for long-term AIMD simulations.
(3) For the compounds investigated in this work fine-tuned MLFFs are transferable. 
This is somewhat surprising, because even though the compounds are similar, they feature distinct chemical environments, in particular the coexistence of the common phosphate anion with the exotic tetrahydroxyphosphonium cation in CPP.

\begin{figure}[ht]
    \centering
    \includegraphics{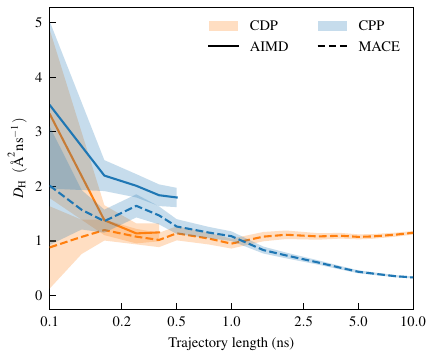}
    \caption{
     Visualization of the convergence behavior of the hydrogen diffusion coefficient with respect to simulation time for CDP  and CPP. 
     AIMD simulations at 513~K and fully fine-tuned mace simulations at 510~K are compared.
     Convergence of the diffusion coefficient is achieved only for timescales above 5~ns, which are not accessible by AIMD simulations.
     The overall diffusion coefficient (represented by solid and dashed lines) is computed as the average of the diffusion coefficients of all individual protons. 
     The shaded area around each line represents the standard deviation calculated for these sets of individual diffusion coefficients.
    }
    \label{fig:plot3}
\end{figure}

We now turn to the final question: Can we obtain "correct" transport properties for solid acids?
Since solid acids are potential membrane materials for fuel cells, the diffusion coefficients are of much more general interest than the RDFs. 
The Grotthuss diffusion mechanism of protons in solid acids consist of an alternating sequence of proton jumps between neighboring anions and rotations of these anions. 
Both processes are crucial and equally important for long-range proton transfer. 
Only a balanced interplay of both contributions allows efficient proton transport. 
Without anion rotation, proton motion would be limited to rattling between two neighboring oxygen atoms. 
Without proton jumps between neighboring anions, no long-range proton transfer could be observed at all, since the centers of mass of anions and cations are fixed due to the crystallinity of the solid acids.
The characteristic timescale of phosphate rotation in these compounds is on the order of several hundred picoseconds \cite{dressler2020c, dressler23}.
These time scales (1) represent the minimum length of an MD simulation for calculating converged diffusion coefficients, (2) are difficult (or impossible) to achieve with AIMD simulations, (3) but are easily accessible with MLFFs.

We have calculated the diffusion coefficients from multi-nanosecond MACE simulations using the fully fine-tuned models (FT@228ps for CDP and FT@233ps for CPP), see details in the Methods section.
For the sake of comparability, we have also performed AIMD simulations for exceptionally long time scales (CDP: 420~ps, CPP: 475~ps).  
The calculation of these trajectories required a significant amount of computational resources. 
For example, we used about $5\cdot 10^{5}$ CPU hours to compute the 475~ps of CPP, which is equivalent to 12 months on a 64-core compute node. 
It is obviously not possible to run such long AIMD simulations in standard workflows for predicting the properties of a large number of materials.
However, the 10~ns MACE trajectories were calculated well within 10 days on a single computing node with an NVIDIA~A100 graphics card - a speedup by a factor of around 700. 

Figure~\ref{fig:plot3} visualizes the diffusion coefficients as a function of trajectory length. 
Starting from the 100~ps AIMD simulation (marking the beginning of the lines in Fig.~\ref{fig:plot3}), the diffusion coefficient computed over about twelve months on a 64-core CPU differs significantly (by about a factor of four) from the converged values obtained with much longer MACE simulations.
Convergence of the diffusion coefficients was achieved after about $5$~ns for both solid acids. 
Such long time frames are obviously inaccessible with any reasonable amount of computational resources using AIMD simulations for the system sizes studied.

The agreement between the diffusion coefficient obtained from very long AIMD simulations and the fully fine-tuned MACE model is convincing. 
However, the direct comparison of these values is problematic due to the limited time scales of the AIMD simulations. 

\section*{Discussion}

The converged diffusion coefficient values for the fully finetuned models, especially for CPP, finally resolve an open question.
On the short AIMD timescales, a smaller diffusion coefficient of CDP compared to CPP was predicted at 510~K, and the diffusion coefficient of CPP did not differ significantly between 410~K and 510~K \cite{dressler23}.
These predictions were in direct contradiction with the conductivity measurements (e.g., see Fig.~10 in Ref.~\citenum{wang22} and Fig.~8 in Ref.~\citenum{wang20}) and indicated a qualitative wrong trend for the activation energy.
If we analyze the diffusion coefficients obtained from the converged MACE simulations, with trajectory lengths longer than 5~ns, we obtain a significantly altered picture, in agreement with the experimental trends for the diffusion coefficients.
While the diffusion coefficients obtained from the long-term AIMD and converged (even longer running) MACE simulations are in good agreement for CDP, they differ by a factor of 4 for CPP.
These deviations are the origin of the qualitatively wrong picture we get from the AIMD simulations.
In contrast, the activation energies for proton diffusion obtained from the MACE simulations (CDP: 0.41~eV, CPP: 0.61~eV) are in good agreement with the experimental activation energies (CDP: 0.40~eV, CPP: 0.65~eV).
Details of the calculation of the activation energies are given in Supplementary~Note~4.

In summary, our work highlights the importance of fine-tuning MLFFs to achieve accurate quantitative descriptions of complex proton transport dynamics in solid acids. 
While the MACE foundation model qualitatively describes all RDFs within this class very well, fine-tuning is essential for precise quantitative agreement with the AIMD reference data.
Within the similar compounds investigated in this work, fine-tuned MLFFs are transferable. 
Using the fine-tuned MACE models, we were finally able to reconcile theoretical predictions with experimental trends for hydrogen diffusion coefficients, especially for CPP.
This underscores the utility of the MLFF MACE in advancing accurate simulations of diffusion phenomena in complex materials, especially solid acids. 
Arguably, these results open the possibility of studying diffusion processes in larger systems at different temperatures and pressures, and potentially indicate that AIMD simulations are now "just" a tool for generating training data for MLFFs.

In conclusion, our work examines the performance of state-of-the-art foundational MLFF models on highly complex solid acids, provides insight into their fine-tunability and transferability, and lays the groundwork for future studies of diffusion phenomena across diverse material systems.

\section*{Methods}

\subsection*{AIMD simulations}

We performed \textit{ab initio} molecular dynamics simulations (Born-Oppenheimer MD scheme) using the CP2K \cite{kuhne2020cp2k} program package to describe the solid acids CDP and CPP.
We utilized the Quickstep module \cite{VandeVondele-2005-ID587} and orbital transformation\cite{VandeVondele-2003-ID584} for faster convergence. 
The electronic structure was calculated with density functional theory utilizing the PBE \cite{Perdew1996} functional.
A basis set of the type DZVP-MOLOPT-SR-GTH \cite{VandeVondele-2007-ID676} and GTH pseudopotentials\cite{Hartwigsen-1998-ID678, Krack-2005-ID586} were applied. 
Furthermore, we used the empirical dispersion correction (D3) to improve the accuracy of the trajectories \cite{Grimme-2010-ID679}.
The temperature was set by a Nos\'{e}-Hoover Chain thermostat \cite{Nose-1984-ID135, Martyna-1992-ID680, Nose-1970-ID681} (NVT ensemble). 
All AIMD simulations were performed with a time step of 0.5~fs.
We have performed AIMD simulations for 420~ps for CDP at 513~K and for 475~ps for CPP also at 513~K.
The simulated systems contained 512 atoms (including 128 H atoms) for CDP and 576 atoms (including 160 H atoms) for CPP.  
The dimensions of the simulation box and the starting configurations of the systems were obtained from crystal structure data from the literature \cite{wang20,yamada04}.

\subsection*{MACE fine-tuning}

Starting from the MACE-MP-0 foundation model, we have created several fine-tuned models for CDP and CPP using the MACE python package (version 0.3.6) \cite{batia2023}. 
To provide reference data for refinement, we recalculated energies and forces using CP2K on different numbers of frames from the AIMD trajectories. Compared to the settings for the AIMD simulation, we only switched off the Grimme dispersion correction, as this correction is also available in the MACE python package and can be used optionally during the production run.
For both compounds (CDP and CPP) we used every 10th snapshot from a $0.01$~ps, $0.1$~ps, and $1$~ps trajectory, resulting in 2, 20, and 200 snapshots for fine-tuning. The resulting models were named by FT@X with X $\in \{0.01\mathrm{ps}, 0.1\mathrm{ps}, 1\mathrm{ps}, 228\mathrm{ps}\}$.
The full fine-tuned models were obtained by retraining the MACE-MP-0 foundation model on every 100th snapshot of a $228$~ps ($233$~ps) trajectory for CDP (CPP).
The training error is reported in Supplementary~Note~2.
We fine-tuned on both energies and forces and used the following hyperparameters for fine-tuning:
Learning rate: 0.01;
Number of epochs: 200;
Training batch size: 5;
Force error weight: 10;
Energy error weight: 0.1.

\subsection*{MACE MD Simulations}

We used the ASE Python package to perform MD simulations using the MACE models for force predictions. 
For the fully fine-tuned models, we performed 10~ns simulation runs of CDP and CPP at 510~K and 610~K to investigate trends in the diffusion coefficients. 
The time step for these simulations was set to 0.5~fs and a Langevin thermostat was used to maintain a constant temperature.

\subsection*{MSD and Diffusion Coefficients}

The mean square displacement $\mathrm{MSD}(\tau)$ is calculated according to 
\begin{eqnarray}
\mathrm{MSD}(\tau) &=& \big\langle \left\langle |\mathbf{R}(t_0+\tau) - \mathbf{R}(t_0)  |^2 \,\, \right\rangle_\mathrm{H}\big\rangle_{t_0}
\end{eqnarray}
Here, $\mathbf{R}$ denotes the position of the hydrogen atoms and the double average $\left\langle\; \left\langle \cdot
\right\rangle_{H}\;\right\rangle_{t_0}$ corresponds to an averaging over all hydrogen atoms $H$ as well as an averaging over all possible starting times $t_0$ within the trajectory. 
For $t_0$, a large number of time points along the total simulation is taken, in order to avoid a bias due to a particular choice of the initial configuration (the "zero" in time) for the actual mean square displacement  $\big\langle |\mathbf{R}(t_0+\tau) - \mathbf{R}(t_0) |^2 \big\rangle_\mathrm{H}$.
The final value of the diffusion coefficient was obtained by a linear fit of the linear part of the mean square displacement between 20 and 100~ps.

\section*{Data availability}
The data supporting the findings of this study have been included as reference material to the reviewers and will be made publicly available upon publication.

\section*{Code availability}
The workflow utilized to produce and post-process the results presented in this study are available from the corresponding author upon reasonable request.

\section*{Acknowledgements}
We thank the staff of the Compute Center of the Technische Universität Ilmenau and especially Mr.~Henning~Schwanbeck for providing an excellent research environment.

\section*{Competing interests}
All authors declare no financial or non-financial competing interests. 

\bibliography{literature.bib}

\end{document}